\documentclass[apj,twocolumn,twocolappendix,floatfix]{openjournal}

\DeclareRobustCommand{\VAN}[3]{#2}
\let\VANthebibliography\thebibliography
\def\thebibliography{\DeclareRobustCommand{\VAN}[3]{##3}\VANthebibliography}

\makeatletter
\def\frontmatter@title@below{\vspace*{0.1in}}
\def\frontmatter@above@affilgroup{\vspace*{0.1in}}
\def\frontmatter@above@affiliation{\vspace*{0pt}}
\makeatother

\usepackage{xcolor}
\usepackage{textgreek}
\usepackage[utf8]{inputenc}
\usepackage[english]{babel}

\usepackage[nolist]{acronym}

\newcommand{\ergcm}[1]{erg~cm$^{-2}$ s$^{-1}$}

\newcommand{\HI}{H{\sc i}}

\newcommand{\Halpha}{H${\alpha}$}

\newcommand{\D}{$^\circ$}

\newcommand{\agn}{NGC\,7213}
\newcommand{\agnold}{Sand-timer}

\def\arcmin{\hbox{$^\prime$}}
\def\arcsec{\hbox{$^{\prime\prime}$}}
\def\kms{km\,s$^{-1}$}

\def\point#1{\hbox{\setbox7=\hbox to0.6em{\hfil.\hfil}%
\setbox8=\hbox to0.5em{\hfil$^{#1}$\hfil}%
\box7\kern-0.5em\box8}}

\def\pointmin#1{\hbox{\setbox2=\hbox to0.8em{\hfil.\hfil}%
\setbox3=\hbox to0.6em{\hfil$^{#1}$\hfil}%
\box2\kern-.7em\box3}}

\def\mmm{\pointmin{\mathrm{m}}\kern.15em}

\usepackage{hyperref}
\hypersetup{
    unicode, 
    colorlinks=true,
    linkcolor=linkcolor,
    citecolor=linkcolor,
    filecolor=linkcolor,
    urlcolor=linkcolor,
}
\usepackage{color,colortbl}
\definecolor{linkcolor}{rgb}{0.0,0.3,0.5}
\usepackage{tensind}
\tensordelimiter{?}
\DeclareGraphicsExtensions{.bmp,.png,.jpg,.pdf}
\usepackage{verbatim}
\usepackage[normalem]{ulem}
\usepackage{orcidlink}
\usepackage{soul}

\graphicspath{ {./figs/} }

\begin{document}

\title{EMU detection of mysterious multiple component radio-continuum emission from nearby spiral Seyfert~I type galaxy `\agnold' (\agn)}

\author{M. D. Filipovi\'c$^{1}$\orcidlink{0000-0002-4990-9288}}
\email{m.filipovic@westernsydney.edu.au}
\author{Z. J. Smeaton$^1$\orcidlink{0009-0009-7061-0553}}
\author{B. S. Koribalski$^{2,1}$\orcidlink{0000-0003-4351-993X}}
\author{R. Kothes$^{3,4}$\orcidlink{0000-0001-5953-0100}}
\author{A. Ahmad$^{1}$\orcidlink{0000-0002-0457-3661}}
\author{R. Z. E. Alsaberi$^{1,5}$\orcidlink{0000-0001-5609-7372}}
\author{C. Bordiu$^{6}$\orcidlink{0000-0002-7703-0692}}
\author{A. C. Bradley$^{1}$\orcidlink{0009-0007-4802-2793}}
\author{E. J. Crawford$^{1}$\orcidlink{0000-0001-5197-1091}}
\author{A. M. Hopkins$^{7}$\orcidlink{0000-0002-6097-2747}}
\author{E. J\"arvel\"a$^{8}$\orcidlink{0000-0001-9194-7168}}
\author{S. Lazarevi\'c$^{1,2,9}$\orcidlink{0000-0001-6109-8548}}
\author{D. Leahy$^{10}$\orcidlink{0000-0002-4814-958X}}
\author{J. L. Payne$^{1}$\orcidlink{0000-0001-6549-2454}}
\author{S. F. ur Rahman$^{11}$\orcidlink{0000-0001-9414-175X}}
\author{G. P. Rowell$^{12}$\orcidlink{0000-0002-9516-1581}}
\author{H. Sano$^{13}$\orcidlink{0000-0003-2062-5692}}
\author{M. Selby$^{14}$\orcidlink{0009-0002-2870-2971}}
\author{S. S. Shabala$^{15}$\orcidlink{0000-0001-5064-0493}}
\author{E. J. Stanton$^{1}$\orcidlink{0009-0004-0615-0779}}
\author{S. Taziaux$^{16,17}$\orcidlink{0009-0001-6908-2433}}
\author{K. Tsuge$^{13}$\orcidlink{0000-0002-2794-4840}}
\author{D. Uro\v{s}evi\'c$^{18}$\orcidlink{0000-0003-0665-0939}}
\author{N. F. H. Tothill$^{1}$\orcidlink{0000-0002-9931-5162}}
\author{J. Th. van Loon$^{19}$\orcidlink{0000-0002-1272-3017}}
\author{V. Velovi\'c$^{1}$\orcidlink{0000-0002-0416-3267}}
\author{G. L. White$^{1}$\orcidlink{0000-0002-4914-6292}}

\affiliation{$^{1}$Western Sydney University, Locked Bag 1797, Penrith South DC, NSW 2751, Australia}
\affiliation{$^{2}$Australia Telescope National Facility, CSIRO, Space and Astronomy, PO Box 76, Epping, NSW 1710, Australia}
\affiliation{$^{3}$Dominion Radio Astrophysical Observatory, Herzberg Astronomy \& Astrophysics, National Research Council Canada, 717 White Lake Road, Kaleden, BC V0H 1K0, Canada}
\affiliation{$^{4}$Department of Physics, University of Alberta, Edmonton, Alberta, T6G 2E1, Canada}
\affiliation{$^{5}$Faculty of Engineering, Gifu University, 1-1 Yanagido, Gifu 501-1193, Japan}
\affiliation{$^{6}$INAF-Osservatorio Astrofisico di Catania, Via Santa Sofía 78, I-95123 Catania, Italy}
\affiliation{$^{7}$School of Mathematical and Physical Sciences, 12 Wally’s Walk, Macquarie University, NSW 2109, Australia}
\affiliation{$^{8}$Department of Physics and Astronomy, Texas Tech University, Lubbock, TX USA}
\affiliation{$^{9}$Astronomical Observatory, Volgina 7, 11060 Belgrade, Serbia}
\affiliation{$^{10}$Department of Physics and Astronomy, University of Calgary, Calgary, Alberta, T2N IN4, Canada}
\affiliation{$^{11}$Institute of Space and Planetary Astrophysics (ISPA), University of Karachi (UoK), Karachi, Pakistan}
\affiliation{$^{12}$School of Physics, Chemistry and Earth Sciences, The University of Adelaide, Adelaide, 5005, Australia}
\affiliation{$^{13}$Faculty of Engineering, Gifu University, 1-1 Yanagido, Gifu 501-1193, Japan}
\affiliation{$^{14}$Observatorio El Sauce, Chile}
\affiliation{$^{15}$School of Natural Sciences, Private Bag 37, University of Tasmania, Hobart, TAS 7001, Australia}
\affiliation{$^{16}$Ruhr University Bochum, Faculty of Physics and Astronomy, Astronomical Institute, Universitätsstraße 150, 44801 Bochum, Germany}
\affiliation{$^{17}$CSIRO Space and Astronomy, PO Box 1130, Bentley WA 6102, Australia}
\affiliation{$^{18}$Department of Astronomy, Faculty of Mathematics, University of Belgrade, Studentski trg 16, 11000 Belgrade, Serbia}
\affiliation{$^{19}$Lennard-Jones Laboratories, Keele University, ST5 5BG, UK}

\begin{abstract}
We present an \ac{ASKAP} \ac{EMU} radio-continuum detection of the nearby spiral Seyfert~I type galaxy \agn\ which we nickname `sand-timer' due to its morphological appearance. We find intriguing, peculiar emission extending from the central \ac{SMBH}, which we refer to as galaxy hyperbubbles and/or \ac{AGN} jets. 
Finally, we find two previously undetected kiloparsec-scale features in the polarised intensity image, which appear most likely also related to the central \ac{SMBH}, as restarted bipolar radio jets, though somewhat obscured in the radio-continuum total intensity image. They resemble the bow shocks often seen in Fanaroff-Riley Class~II (FR-II) radio galaxies, and so we propose they most likely represent an early episode of the `\agnold's' \ac{AGN} jet activity. 
An alternative scenario is that the core of \agn\ may be an \ac{SMBBH} undergoing a \ac{TDE}, potentially causing the peculiar radio-continuum emission. 
\end{abstract}



\section{INTRODUCTION}
\label{intro}

\begin{acronym}[AWGN]
\acro{2MASS}{Two Micron All Sky Survey}
\acro{2MASX}{Two Micron All Sky Survey Extended Source Catalogue}
\acro{30Dor}[30 Dor]{30~Doradus}
\acro{AeReS}{Aegean Residual}
\acro{AGN}{active galactic nucleus}
\acro{ALMA}{Atacama Large Millimeter Array}
\acro{ARC}{Australian Research Council}
\acro{ATCA}{Australia Telescope Compact Array}
\acro{ATESP}{Australia Telescope ESO Slice Project}
\acro{ATNF}{Australia Telescope National Facility}
\acro{ATOA}{Australia Telescope Online Archive}
\acro{AT20G}{Australia Telescope 20 GHz Survey}
\acro{ASKAP}{Australian Square Kilometre Array Pathfinder}
\acro{BETA}{Boolardy Engineering Test Array}
\acro{BGG}{Brightest Galaxy Group}
\acro{BL}[BL Lac]{BL Lacertae Objects}
\acro{CASS}{CSIRO Astronomy and Space Science}
\acro{CABB}{Compact Array Broadband Back-end}
\acro{CC}{core-collapse}
\acro{CHII}[{\sc CHii}]{Compact \textsc{Hii}}
\acro{CIE}{collisional ionisation equilibrium}
\acro{CMB}{Cosmic Microwave Background}
\acro{CSIRO}{Australian Commonwealth Scientific and Industrial Research Organisation}
\acro{CSS}{Compact Steep Spectrum}
\acro{DESI}{Dark Energy Spectroscopic Instrument}
\acro{DS9}[\textsc{DS9}]{\textsc{SAOImage DS9}}
\acro{DSS}{Digital Sky Survey}
\acro{DSA}{diffuse shock acceleration}
\acro{EBHIS}{Effelsberg--Bonn H\,{\sc i} Survey}
\acro{EM}{Electromagnetic}
\acro{EMU}{Evolutionary Map of the Universe}
\acro{ev}[eV]{electronvolt\acroextra{: 1 eV $\approx 1.6 \times 10^{-19}$ J}}
\acro{eROSITA}{extended R\"Ontgen Survey with an Imaging Telescope Array}
\acro{Fermi-LAT}{Fermi Large Area Telescope}
\acro{FIR}{Far Infrared}
\acro{FITS}[\textsc{Fits}]{Flexible Image Transport System}
\acro{FRB}{Fast Radio Bursts}
\acro{FRED}{fast-rise, exponential-decay}
\acro{FSRQ}{Flat Spectrum Radio Quasars}
\acro{FWHM}{Full Width at Half-Maximum}
\acro{GASS}{Galactic All-Sky Survey}
\acro{GLEAM}{GaLactic and Extragalactic All-sky MWA}
\acro{GPS}{Gigahertz Peak Spectrum}
\acro{HEMT}{High Electron Mobility Transistor}
\acro{HERITAGE}{HERschel Inventory of The Agents of Galaxy Evolution}
\acro{H.E.S.S.}{High Energy Stereoscopic System}
\acro{HFP}[HFP]{High-Frequency Peaker}
\acro{HPBW}{Half Power Beam Width} 
\acro{HzRGs}{High Redshift Radio Galaxies}
\acro{pHFP}[pHFP]{Potential High Frequency Peaker}
\acro{HI}[H{\sc i}]{Neutral Atomic Hydrogen} 
\acro{HST}{\textit{Hubble Space Telescope}}
\acro{ICM}{intracluster medium}
\acro{IAU}{International Astronomical Union}
\acro{IFRSs}{Infrared Faint Radio Sources}
\acro{ISM}{interstellar medium}
\acro{IFRS}{Infrared Faint Radio Source}
\acro{IR}{Infrared}
\acro{JY}[Jy]{Jansky\acroextra{, 1 Jy = $10^{-26} \times \mathrm{W~ m}^{-2}~\mathrm{Hz}^{-1}$}} 
\acro{LLAGN}{low-luminosity active galactic nuclei}
\acro{LINER}{Low-Ionisation Narrow Emission-Line Region}
\acro{LLS}{Largest Linear Size}
\acro{LFAA}{Low-Frequency Aperture Array}
\acro{LMC}{Large Magellanic Cloud}
\acro{LSO}[LSOs]{Large Scale Objects}
\acro{MACHO}{Massive Astrophysical Compact Halo Objects}
\acro{MAGMA}{Magellanic Mopra Assessment}
\acro{MC}[MC]{Magellanic Cloud}
\acro{MS}[MS]{Magellanic Stream} 
\acro{MB}[MB]{Magellanic Bridge} 
\acro{MCELS}{Magellanic Cloud Emission Line Survey}
\acro{MW}{Milky Way}
\acro{MIPS}{Multiband Imaging Photometer}
\acro{MIRIAD}[\textsc{Miriad}]{Multichannel Image Reconstruction, Image Analysis and Display}
\acro{MIT}{Massachusetts Institute of Technology}
\acro{MM}{mixed-morphology}
\acro{MOST}{Molonglo Observatory Synthesis Telescope}
\acro{MQS}{Magellanic Quasars Survey}
\acro{MRC}{Molonglo Reference Catalogue of Radio Sources}
\acro{MT-MFS}{Multi-Term Multi-Frequency Synthesis}
\acro{MWA}{Murchison Widefield Array}
\acro{NED}{NASA/IPAC Extragalactic Database}
\acro{NRAO}{National Radio Astronomy Observatory}
\acro{NVSS}{NRAO VLA Sky Survey}
\acro{OPAL}{Online Proposal Applications \& Links}
\acrodefplural{ORC}{Odd Radio Circles}
\acro{ORC}{Odd Radio Circle}
\acro{OVV}{Optically Violent Variable Quasars}
\acro{PACS}{Photodetector Array Camera and Spectrometer}
\acro{PAF}{Phased Array Feed}
\acro{pc}{parsec\acroextra{: 1 pc $\simeq 3.09 \times 10^{16}$ m}}
\acro{PMN}{Parkes-MIT-NRAO}
\acro{PN}{Planetary Nebula}
\acro{PWN}{pulsar wind nebula}
\acrodefplural{PWN}[PWNe]{pulsar wind nebulae}
\acro{RN}{reflection nebula}
\acrodefplural{RN}[RNe]{reflection nebulae}
\acro{QSO}{Quasi-Stellar Object}
\acro{RA}{Right Ascension}
\acro{RFI}{Radio-Frequency Interference}
\acro{RIAF}{radiatively inefficient accretion flow}
\acro{RMS}{Root Mean Squared}
\acro{SAGE}{Surveying the Agents of a Galaxy’s Evolution}
\acro{SARAO}{South African Radio Astronomy Observatory}
\acro{SDSS}{Sloan Digital Sky Survey}
\acro{SED}{spectral energy distribution}
\acro{SI}[$\alpha$]{Spectral Index\acroextra{, $S \propto \nu^\alpha$}}
\acro{SHS}{SuperCOSMOS H$\alpha$ Survey}
\acro{SKA}{Square Kilometre Array}
\acro{SMBH}{supermassive black hole}
\acro{SMBBH}{super massive binary black hole}
\acro{POSSUM}{Polarisation Sky Survey of the Universe's Magnetism}
\acro{CASDA}{CSIRO ASKAP Science Data Archive}
\acro{SMC}{Small Magellanic Cloud}
\acro{SN}{Supernova}
\acro{SPIRE}{Spectral and Photometric Imaging Receiver}
\acro{SUMSS}{Sydney University Molonglo Sky Survey}
\acro{TOPCAT}[\textsc{Topcat}]{Tool for OPerations on Catalogues And Tables}
\acro{TDE}{tidal disruption event}
\acro{USS}{Ultra Steep Spectrum}
\acro{YSOs}{young stellar objects}
\acro{WBAC}{Wide-Band Analogue Correlator}
\acro{WIFES}[WiFeS]{Wide-Field Spectrograph}
\acro{WISE}{Wide-Field Infrared Survey Explorer}
\acro{VLBI}{Very Long Baseline Interferometry} 
\acro{VLSR}[\textbf{$v_{lsr}$}]{Velocity in the Line of Sight}
\acro{SMASH}{Survey of the MAgellanic Stellar History}
\acro{SNR}{supernova remnant}
\acro{SD}{single-degenerate}
\acro{SUMSS}{Sydney University Molonglo Sky Survey}
\acro{TDE}{tidal disruption event}
\acro{CARTA}{Cube Analysis and Rendering Tool for Astronomy}
\acro{NEI}{non-equilibrium ionisation}
\acro{UV}{ultra-violet}
\acro{FUSE}{Far Ultraviolet Spectroscopic Explorer}
\acro{RM}{rotation measure}
\acro{H.E.S.S.}{High Energy Spectroscopic System}
\acro{CTA}{Cherenkov Telescope Array}
\acro{WR}{Wolf-Rayet}
\end{acronym}

\acresetall

\agn\footnote{Here nicknamed `\agnold' due to its morphological appearance as seen in the radio-continuum.} (ESO~288-~G~043) is a nearby face-on spiral (SA(s)a) galaxy \citep{1991rc3..book.....D}, though with an inclination angle of $i$=21\D$^{+7}_{-13}$ \citep{2015MNRAS.454.1157P}. It is a member of the poor Indo-Pavus group \citep{1975gaun.book..557D}, with a measured redshift of $z$=0.005839~\citep{2006MNRAS.371.1855W}. However, the distance to \agn\ is best obtained from the \HI\ systemic velocity \citep[1762$\pm$15\,\kms;][]{2004AJ....128...16K} of D$\sim$25\,Mpc, which is used in this paper. The source seems rather isolated, not in a rich cluster environment, which may explain why it has avoided recent strong interactions.

\begin{figure*}[ht!]
    \centering
    \includegraphics[width=1\linewidth]{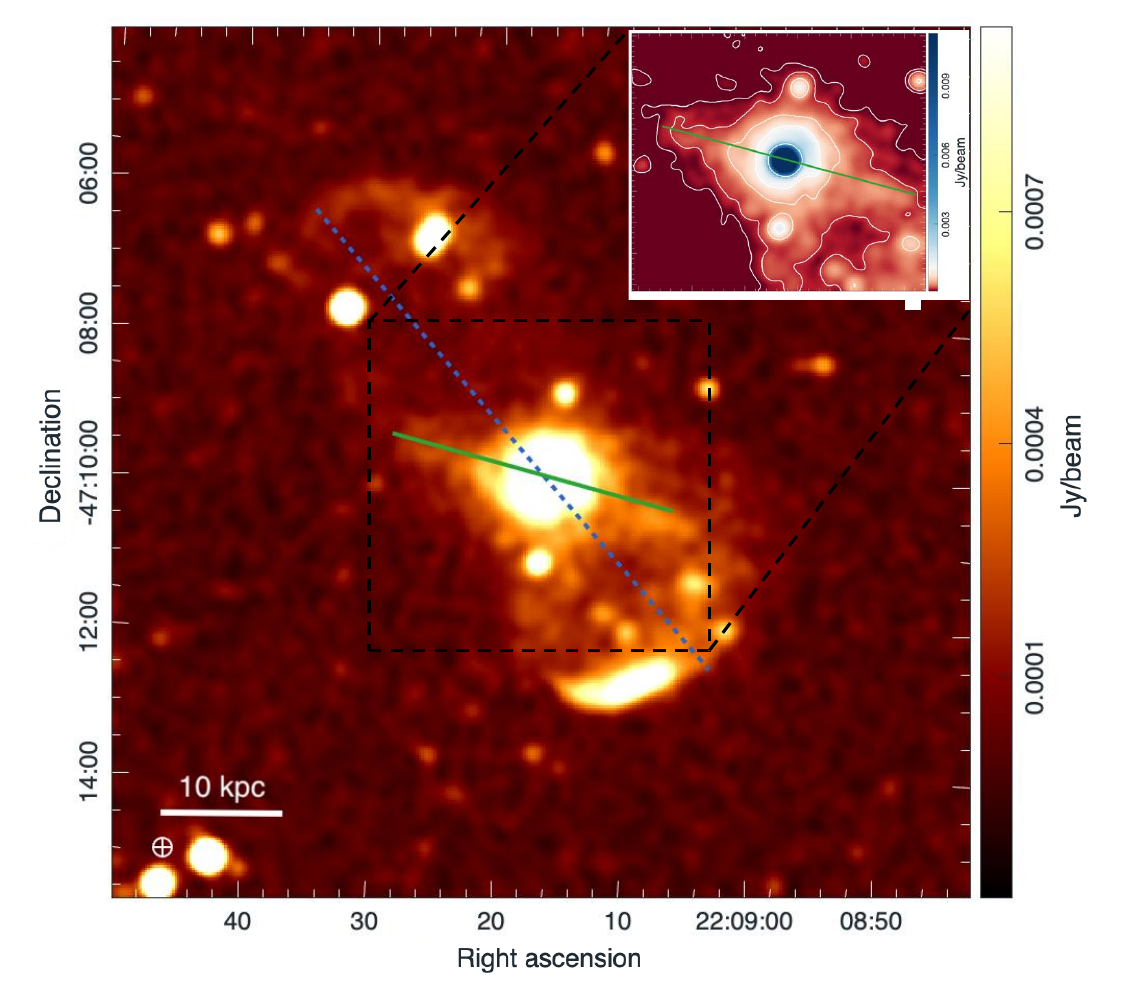}
    \caption{Radio-continuum image of \agn\ with \ac{EMU} data at 944\,MHz. The image has a measured local \ac{RMS} noise level of $\sim$25\,$\mu$Jy\,beam$^{-1}$ and a beam size of 15\arcsec$\times$15\arcsec\ shown in the bottom left corner. The image intensity is linearly scaled with the colour bar shown on the right. The axis of the larger lobes is shown by a dashed blue line and has a position angle of 40$^\circ$, and the smaller possible jet axis is marked with the solid green line with a position angle of 75$^\circ$ (also see Figure~\ref{fig:polarisationWOptical2} (bottom)). Both position angles are measured counter-clockwise from North. The inset in the top-right corner shows a zoomed-in image of the core and a possible jet structure. This image uses a logarithmic scale (with $\alpha$=100) and a different colour map to emphasise the fainter jet structure next to the brighter core. The green line showing the possible jet axis is also shown in the inset image. The inset shows contour levels from the radio-continuum image at 0.1, 0.2, 0.5, 1.0, and 5.0\,mJy\,beam$^{-1}$.}
    \label{fig:EMU}
\end{figure*}


Sand-timer's nature as an \ac{AGN} is widely debated in the literature as it was initially classified as a Seyfert~I by \citet{1979ApJ...227L.121P} and \citet{2017ApJS..232...11T}, and was later identified as a \ac{LINER} due to the narrow-line emission in its optical spectrum~\citep{1984ApJ...285..475H, 2005MNRAS.356..727S}. \agn\ is reported as having a black hole mass of $\sim10^8M_\odot$~\citep{2002ApJ...579..530W,2014MNRAS.438.3322S} and a bolometric luminosity of $1-2\times10^{43}$\,erg\,s$^{-1}$~\citep{2013MNRAS.429.3439E,2020A&A...641A.151S}. \citet{2020A&A...641A.151S} also calculated an Eddington ratio of $\sim9\times10^{-4}$; lower than expected for a typical Seyfert~I galaxy. This low Eddington ratio suggests radiatively inefficient accretion. The study of \citet{2011MNRAS.411..402B} indicated that \agn\ is one of the most luminous \ac{LLAGN}. 
\agn\ has also been analysed at radio-continuum frequencies and was classified as a compact radio source with a radio power larger than expected for a typical Seyfert galaxy~\citep{1998ApJ...497..133B}, but not sufficient to be classified as a radio-loud \ac{AGN}~\citep{2005MNRAS.356..734B}. This larger-than-expected radio power may be attributed to the low Eddington ratio, as lower Eddington ratios are observed to produce more radio-loud AGN~\citep{2002ApJ...564..120H}. \citet{2011MNRAS.411..402B}, based on \ac{ATCA} and {\it Rossi X-ray Timing Explorer} (RXTE) observations, suggested that \agn\ can be considered to be comparable to a hard-state black hole X-ray binary because its central \ac{SMBH} shows substantial variability at both radio and X-ray frequencies. 

Unlike many spirals, \agn\ is deficient in cold gas and shows suppressed star formation. The lack of molecular gas near the centre may explain why the \ac{AGN} dominates while the galaxy is otherwise relatively quiescent. The \ac{ALMA} study of \agn\ also analysed the molecular gas content and found a faint, unobscured \ac{AGN} that does not significantly impact the star formation properties of its host galaxy~\citep{2020A&A...641A.151S}. 

In 2015, a bright supernova called ASASSN-15us \citep{2015ATel.8474....1B} was discovered in \agn\ by the All Sky Automated Survey for SuperNovae (ASAS-SN). Unfortunately, no detection has been made at the radio-continuum frequencies.

Previous radio-continuum observations of \agn\ detected only the central, radio-bright core but lacked the sensitivity to detect the faint structures extending from the core. To add to our collection of discoveries in the radio-continuum regime, we present the \ac{EMU} view of \agn. The new radio-continuum observations that we present here offer a detailed view into its morphology, revealing the presence of faint and heretofore unseen radio structures extending from the central core (see Figure~\ref{fig:EMU}).

\begin{figure}[ht]
    \centering
    \includegraphics[width=1.0\linewidth]{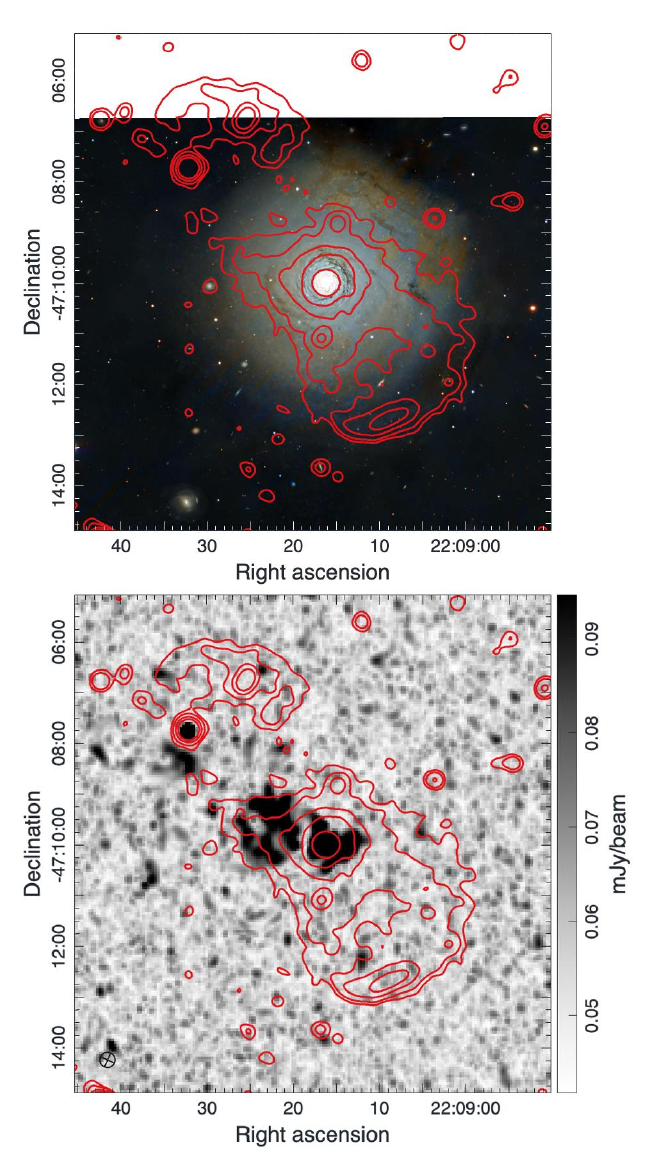}
    \caption{Optical RGB, continuum and polarised intensity image of NGC\,7213. 
    {\bf Top:} Optical RGB of NGC\,7213 (see Section~\ref{Data:Optical}). 
    {\bf Bottom:} Polarised intensity image with an \ac{RMS} noise level of 10\,$\mu$Jy\,beam$^{-1}$. The image is linearly scaled with the colour bar shown on the right side, and the contours are from the 944\,MHz radio-continuum image at the levels of 0.1, 0.2, 0.5, 1.0 and 5.0\,mJy\,beam$^{-1}$. The image is convolved to a beam size of 18\arcsec$\times$18\arcsec, as shown in the bottom-left corner.
}
    \label{fig:polarisationWOptical2}
\end{figure}

\begin{figure*}[ht]
    \centering
    \includegraphics[width=1.0\linewidth]{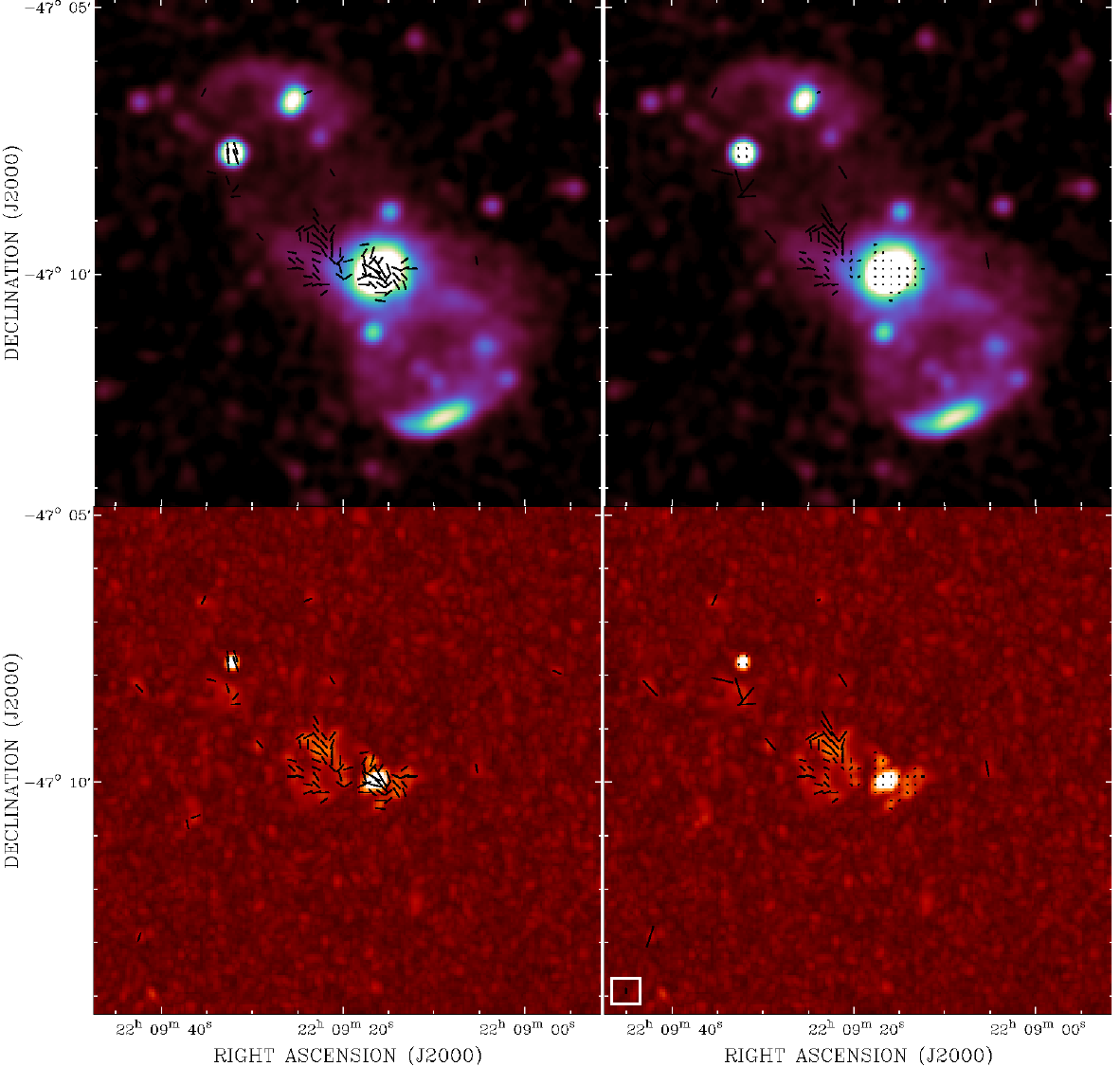}
    \caption{ 
    {\bf Top left:} Radio-continuum image with B-vectors plotted on top.
    {\bf Bottom left:} Polarised intensity image with B-vectors plotted on top. 
    {\bf Top right:} Radio-continuum image with fractional polarisation B-vectors plotted on top.
    {\bf Bottom right:} Polarised intensity image with fractional polarisation B-vectors plotted on top. For the fractional polarisation vectors (right column), the B-vectors have been scaled to be proportional to the polarised intensity values. In the lower-left corner of the bottom-right image (black vertical line inside the white square), there is a vertical vector indicating 10\% polarisation.
}
    \label{fig:polarisationWOptical3}
\end{figure*}

\section{DATA}
\label{data}

\subsection{\ac{ASKAP}}
\label{Data:ASKAP}

\subsubsection{EMU}

As part of the large-scale \ac{ASKAP}-\ac{EMU} project (AS201) \citep{Norris2021,2025PASA...42...71H}, this part of the radio sky was observed on 22$^{\rm nd}$~August~2025 for 10~hours with a complete set of 36 \ac{ASKAP} antennas at the central frequency of 943.5\,MHz and bandwidth of 288~MHz. All data are available in the CSIRO \ac{ASKAP} Science Data Archive (CASDA\footnote{\url{https://research.csiro.au/casda}}). The observations of this object are on tile EMU\_2214$-$46, corresponding to \ac{ASKAP} scheduling block SB76123. 

We processed the \ac{EMU} data ourselves using 12 beams and the ASKAPsoft pipeline, including multi-frequency synthesis imaging, multi-scale clean and self-calibration \citep{askapsoft_2019ascl.soft12003G}. The final 944\,MHz Stokes~$I$ radio image is shown in Figure~\ref{fig:EMU}, and we estimate a local \ac{RMS} noise sensitivity of $\sim$25\,$\mu$Jy\,beam$^{-1}$. We also analyse the available Stokes~$V$ \ac{EMU} data, but find no corresponding emission.

\begin{figure*}[ht!]
    \centering
    \includegraphics[width=0.95\linewidth]{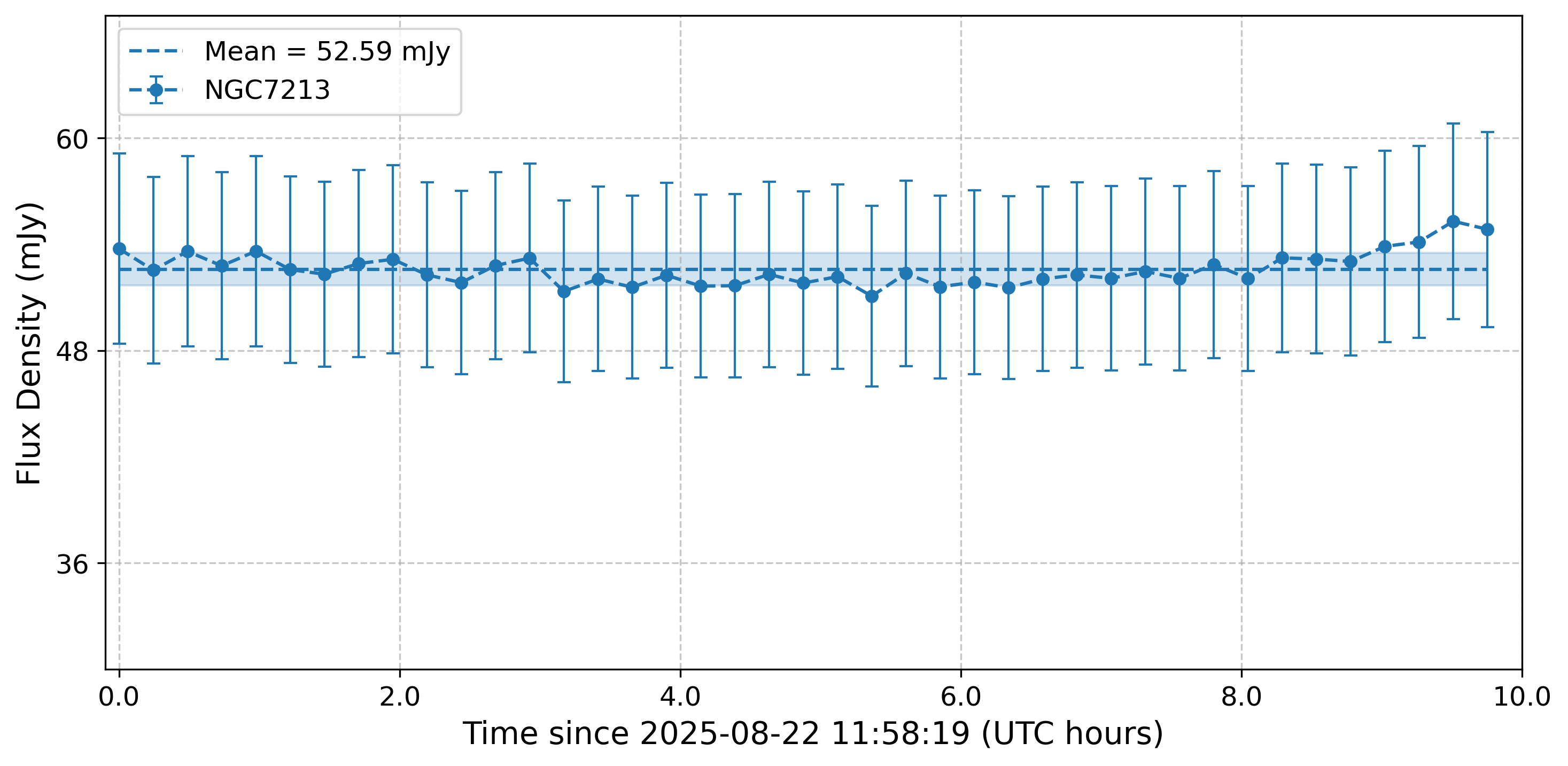}
    \caption{Flux density light curve of the central source on the timescale of 15~minutes. The 1$\sigma$ standard deviation of 3.70\,mJy/beam from the mean flux density (dashed line) is represented by the shaded region.} 
    \label{fig:lightcurve}
\end{figure*}

In order to assess the variability of the central source, we created Stokes~{\it I} images every 15~minutes ($\sim$41 snapshot images) over the 10-hour-long observations. 
The flux density of the central core was measured by integrating over tight elliptical apertures (1.5$\times$ the beam size of each snapshot image) centred on the source's peak position. 
Some of the \ac{ASKAP} images have a flipped RA axis, so the beam position angle from the FITS header must be rotated by +90\D\ to match the true pixel‑space orientation. This is required, as otherwise the aperture would become increasingly misaligned as the beam elongated toward the end of the track (e.g after snapshot T36), causing a significant part of the on-source pixels to fall outside the measurement region and producing an artificial decline. After applying the correct +90\D\ transformation and extracting the flux, the target source is stable across the full observation, and nearby sources with similar flux densities show the same stable behaviour, meaning NGC\,7213 is not significantly varying during the 10-hour track.
We estimated the local \ac{RMS} noise by taking the standard deviation within a larger aperture while excluding the on-source pixels, finding a typical local \ac{RMS} noise of $\sim$170\,$\mu$Jy\,beam$^{-1}$. 
We then construct a light curve using 41 flux density vs. time measurements as shown in Figure~\ref{fig:lightcurve}. 
We adopted a 10\% fractional uncertainty \citep[][]{Filipovic2022} on the measured flux density from each snapshot image, which is plotted as error bars.

\subsubsection{POSSUM}

A full description of the \ac{POSSUM} data processing pipeline can be found in \citet{2025PASA...42...91G}. This pipeline handles data calibration, imaging, and the production of images and final data cubes. The instrumental bandpass and on-axis instrumental polarisation calibration are derived using the primary calibrator PKS~B1934--638. About every 24~hours, this calibrator is observed for a few minutes at the centre of each beam, and the resulting calibration is then applied to the science data during processing \citep{2025PASA...42...91G}. Models of the \ac{ASKAP} primary beams are derived from holographic measurements of the unresolved, unpolarised calibrator PKS~B0408--65 \citep{2025PASA...42...91G}. These observations allow for the modelling and correction of off-axis instrumental polarisation effects. These instrumental polarisation correction procedures produce data with residual leakage of about 0.2\% from Stokes $I$ into $Q$, $U$, and $V$ \citep{2025PASA...42...91G}.
Released \ac{POSSUM} data consist of Stokes~$Q$ and $U$ frequency data cubes with 1\,MHz channels. The polarised intensity (PI) map presented here (see Figures~\ref{fig:polarisationWOptical2} and \ref{fig:polarisationWOptical3}, bottom), by de-rotating the $Q$ and $U$ data in each frequency channel for each \ac{RM} and using the data at the peak of the Faraday depth function for each pixel, as described in \citet{Ball2023}. The final polarised intensity map has a resolution of 18\arcsec. Note that the length of the B-vectors is proportional to polarised intensity.

\subsection{\ac{ATCA}}
\label{sec:ATCA_DATA}

We used archival\footnote{\ac{ATOA}, hosted by the \ac{ATNF}: \url{https://atoa.atnf.csiro.au}} \ac{ATCA} data (project code: CX579). Further information on the \ac{ATCA} data, including details of the observations and data reduction, is provided in \citet{2026A&A...710A.217B}.

\subsection{\ac{MWA}}
\label{Data:MWA}

We also used \ac{MWA} radio-continuum data from the \ac{GLEAM} survey~\citep{2015PASA...32...25W, 2017MNRAS.464.1146H, 2019PASA...36...47H, 2019PASA...36...48H}. These data consist of four images centred at 88, 118, 155, and 200\,MHz. The central frequency, bandwidth, and beam size differ across images, and the values are presented in Table~\ref{tab:MWA}. The local \ac{RMS} value of each image was also measured from a background region near the source, and the estimated value is also reported in Table~\ref{tab:MWA}.

\begin{table}[ht]
    \centering
    \scriptsize
        \caption{Details of MWA data used in this study.} \vspace{0.5em}
    \begin{tabular}{ccccc} \hline
        $\nu$ & Bandwidth & \ac{RMS} & Beam Size & PA\\
        (MHz) & (MHz) & (mJy beam$^{-1}$) & ($\arcsec\times\arcsec$) & ($^\circ$) \\ \hline
        88 & 30 & 33 & 297$\times$290 & $-$32.0 \\
        118 & 30 & 11 & 213$\times$209 & $-$12.5 \\
        155 & 30 & 6 & 164$\times$158 & $-$1.6 \\
        200 & 60 & 8 & 137$\times$131 & $-$10.8 \\
        \hline\hline
    \end{tabular}
    \label{tab:MWA}
\end{table}

\subsection{Optical}

\label{Data:Optical}

We observed \agn\ on 15$^{\rm th}$ and 17$^{\rm th}$~September~2025 using optical telescopes at the Observatorio El Sauce, Chile (\url{https://obstech.cl/}). We used the Luminance Telescope PlaneWave CDK 1000 FL 6000\,mm (\url{https://planewave.com/products/cdk1000/)} with the Moravian C5~100M camera. 
Filters are Chroma, luminance, and for colours are R (red), G (green) and B (blue).

To process the \agn\ images, we framed their offset due to a very bright nearby foreground star that was washing across the frame. With the field derotation of the CDK~1000, multiple spikes are created from bright stars that can “shower” the target. The final image is shown in Figure~\ref{fig:polarisationWOptical2} (top panel).

\section{RESULTS}
\label{results}

The radio-continuum emission is composed of two bubble-like, elliptical features (expanding from the central \ac{SMBH} of \agn) that extend symmetrically along the north-east to south-west axis (see Figure~\ref{fig:EMU}), which we initially call `galaxy hyperbubbles'. These appear morphologically similar to the galaxy-scale structures which have been observed in other galaxies, e.g. NGC 3079~\citep{1983ApJ...273L..11D,1998A&A...340..351P} and the nearby \ac{LMC}'s 30~Dor~C region~\citep{2015A&A...573A..73K,2017ApJ...843...61S,2019A&A...621A.138K,2021ApJ...918...36Y}. 
These regions and emissions are noted to exhibit predominantly thermal bremsstrahlung emission and have been described using the terms `bubbles', `super-bubbles', and `thermal hyperbubbles', but no consistent naming convention has been defined. We adopt the term `hyperbubbles' throughout this paper to refer to the structures seen in \agn.
Also, in Mrk\,6 \citep{2006ApJ...652..177K} and NGC\,6764 \citep{2010ApJ...723..580K}, precessing jets were invoked to explain the radio structures. NGC\,4235 has been suggested to show episodic \ac{AGN} activity \citep{2016MNRAS.459.1310K}, while NGC\,3079 shows `rings' and other features \citep{2019ApJ...883..189S} similar to \agn. Finally, we note that in NGC\,3516, shells and hotspots are quite similar \citep{2025ApJ...989...40G} to \agn\ but somewhat smaller. In general, the lobes in \agn\ are on average a bit larger than the above examples, where typical lobe sizes might be closer to 10--20\,kpc in Seyferts.

The south-western extension of \agn\ appears brighter in the radio-continuum emission than the north-eastern one, and as a result, it shows a more definitive edge structure, particularly along the brighter southern rim. There is a strong associated total polarisation detected (see Figure~\ref{fig:polarisationWOptical2}). This polarisation could be due to a magnetised outflow, for example, as seen in another similar system~\citet[ESO 130-G012][]{2026PASA...43...22K}, or from precessing \ac{AGN} jets which would originate from a central \ac{SMBBH}(see Section~\ref{discussion}).

We also note a strong radio source along the north-east feature, towards the north end, which is in line with the diffuse radio-continuum emission (see Figure~\ref {fig:EMU}). It is possible that this northern emission is associated with this point source instead of the larger \agn\ structure. The source is located at coordinates RA(J2000)\,=\,22$^{\rm h}$09$^{\rm m}$25.5$^{\rm s}$, Dec(J2000)\,=\,--47$^{\circ}$06\arcmin45\farcs1. We searched SIMBAD and NED at these coordinates but found no listed sources. We found a nearby source, WISEA~J22092497--470612.9, which is listed as a galaxy, but it appears as a distinct source separated by $\sim$30\arcsec. It is possible that this source is an unclassified \ac{AGN}, and the northern emission is instead associated with it, making the larger structure a superposition of multiple objects rather than entirely associated with \agn. We note that this northern radio-continuum emission is not detected at any other frequency, including optical, IR, or X-ray. However, because of the lack of information about this source and the apparent symmetry of the structures around the \agn\ core, we assume the emission is more likely associated with the larger structure and maintain this assumption throughout our analysis.

\subsection{\agn\ Radio Flux Density}
\label{sec:flux}

\subsubsection{ASKAP}
\label{Results:ASKAP}

The central core appears as a compact source in the \ac{EMU} image with the larger structure surrounding it. We fit a region around both the total structure and the inner point source using the \ac{CARTA} visualisation software. We measure an angular size of 8\farcm2$\times$4\farcm0, corresponding to a physical size of 56.7$\times$27.7\,kpc for the entire structure. The north-eastern galaxy hyperbubble is slightly longer than the south-western one, and we measure lengths of 4\farcm7 for the north-eastern hyperbubble and 3\farcm0 for the south-western hyperbubble, corresponding to physical sizes of 32.5 and 20.7\,kpc. We measure the axis of these large-scale lobe structures as having a position angle of 40$^\circ$ (see Figure~\ref{fig:EMU}). The observed bubbles appear outside the host galaxy and thus must have carved their way through the inner galaxy, possibly along the galaxy's plane. Therefore, they may have advanced at different speeds or encountered different densities of intervening material. This would, in turn, affect the relative velocities of the escaping jets and so may result in the observed size discrepancy. Alternatively, this size discrepancy may be due to a viewing angle rather than a physical property; however, this is unlikely given the large distance ($\sim$25\,Mpc) to \agn.

We measure integrated flux densities of $S_{944\,\text{MHz}}\,=\,150.6\pm15.1$\,mJy for the entire structure (see Table~\ref{tab:fluxes}) and $S_{944\,\text{MHz}}\,=\,55.8\pm5.6$\,mJy for the core. We subtract the core flux density from the total structure flux density to measure the diffuse component, and estimate a flux density of $S_{944\,\text{MHz}}\,=\,94.8\pm16.1$\,mJy. The uncertainties for the individual measurements were set to 10\%, and the uncertainty of the diffuse component was estimated by taking the individual uncertainties in quadrature.

Also, as explained in Section~\ref{Data:ASKAP}, we measured the flux density of the central core of \agn\ every 15 minutes over the \ac{EMU}'s 10-hour observations (Figure~\ref{fig:lightcurve}). We find no significant variation in the 10-hour observation.

\subsubsection{MWA}
\label{Results:MWA}

\agn\ is also visible in the \ac{MWA} images, but the entire structure appears as a point source due to the lower resolution. The maximum value given in \ac{CARTA} was larger than the integrated value, which can occur for partially resolved sources. Since it appears as a point source, we took the maximum value in each \ac{MWA} band (88, 118, 155, and 200\,MHz) as the flux density (see Table~\ref{tab:fluxes}). We also find several previous radio-continuum observations at different frequencies, which are also reported in Table~\ref{tab:fluxes} with their corresponding references.

 \subsubsection{ATCA}
\label{sec:ATCAflux}

Finally, we used the \ac{ATCA} data \citep{2026A&A...710A.217B} to search for radio-continuum emission in \agn. Due to a short observing time, only the central core of \agn\ was detected, and, similarly to \citet[][see their figure~1]{2011MNRAS.411..402B}, the peak flux density was measured over a period of eight months 
at both wavelengths (6 and 3\,cm). As evident in \citet{2026A&A...710A.217B} measurements, 
there is a distinctive/dramatic central core variability that varies by $\sim$72\% in its peak flux density activity on a short-period (two weeks) scale, and a more detailed comparative analysis is provided in \citet{2026A&A...710A.217B}.

\subsection{\agn\ Radio Spectral Index and Flux-Density Variability}
\label{sec:spectra}

The radio spectral index ($\alpha$) is defined by $S \propto \nu^\alpha$, where $S$ is the flux density and $\nu$ is the observing frequency. We first explore the \ac{ATCA} measurements. We find that the spectral index is inverted and varies with the peak flux density \citep{2026A&A...710A.217B}, suggesting a complex process in the \agn\ core.

To characterise the spectral index and illustrate the degree of flux-density variability across epochs, we attempt a linear fit to the measurements spanning a broad frequency range and large epoch difference (see Table~\ref{tab:fluxes}), using the \textsc{curve\_fit} function in the \textsc{scipy} Python library. The measurements are non-simultaneous; therefore, this fit should not be interpreted as a full characterisation of the source's intrinsic radio spectrum if it is indeed variable on a longer (years) time-scale basis. The resulting spectral index of $\alpha\sim-0.02\pm0.04$ is nominally flat, which is broadly consistent with \citet{2007ApJS..171...61H}, who found $\alpha=0.192$ based only on three available frequencies. The fit is relatively poor, as reflected in a reduced $\chi^2$ value of 20.70, indicating substantial scatter. This may be driven by intrinsic source variability, a scenario also supported by the \ac{ATCA} data and by \citet{2011MNRAS.411..402B}.

To better characterise the spectral properties at a single epoch, we also fit the \ac{MWA} data alone, since all measurements were taken during the same epoch. These measurements yield a steep spectral index value of $\alpha=-0.91\pm0.01$ (see Figure~\ref{fig:specindex}). However, these points sample only the low-frequency end of the radio spectrum and may not represent the broadband spectral behaviour, even at a single epoch. This observed steepness may reflect the low-frequency \ac{MWA} measurements, as an \ac{AGN} core is expected to show a flat spectrum, so the low-frequency end would be dominated by potential synchrotron-emitting jets or lobes (hyperbubbles). A flat overall spectral index is not unexpected for an active \ac{AGN} core due to efficient particle acceleration, and the steep \ac{MWA} index may reflect low-frequency spectral curvature or absorption processes at a single epoch. The reported uncertainties are the default standard errors from \textsc{curve fit} and therefore statistical errors that almost certainly underestimate the true uncertainty.

Long-term X-ray variability \citep{2018MNRAS.475.1190Y} shows an unusual decades-long \ac{FRED} flare in the X-ray light curve -- the nucleus brightened rapidly then declined exponentially over $\sim$20~years, a striking variability behaviour for most \ac{AGN}, but not unusual for low-luminosity \ac{AGN}. This variability can be even more prominent in low-luminosity \ac{AGN} with lower Eddington ratios, as \agn\ has, which may explain the increased variability. Conversely, this may also provide clues about changes in accretion rate and structure of the \agn\ core.
However, most recently, \cite{2025ApJ...994L..13K} presented new X-ray observations showing that the X-ray flux is much brighter now (in 2025), almost back to its historically brightest level, which was in the 1980s. Therefore, the long X-ray decay is only valid until $\sim$2021. The current average X-ray flux in the 2-10\,keV band is $\sim5\times10^{-11}$\,\ergcm.. While this re-brightening does not affect the large-scale feature, the potential core activity is uncertain and warrants further investigation.

If the source is indeed variable (on a longer (years) time-scale basis), then this makes it difficult to estimate spectral indices, particularly when data from different frequencies are obtained at different times. To account for this, we also estimate a spectral index using only the \ac{EMU} observations, specifically the Taylor~1 and Taylor~0 images. Within the \ac{MT-MFS} framework~\citep{2011A&A...532A..71R}, the spectral index can be estimated as $\alpha\approx{\rm T1}/{\rm T0}$, where T0 and T1 are the zeroth- and first-order Taylor coefficients of the frequency-dependent sky brightness. We note that this approximation assumes the spectrum is well described by a simple power law across the observed band and is subject to inherent limitations given the 288\,MHz bandwidth. Additionally, correlated errors between Taylor terms arising from deconvolution coupling are a known systematic issue, particularly for faint extended structures near bright, compact sources, so the reported indices should be treated as indicative rather than precise measurements.

\begin{figure*}[ht!]
    \centering
    \includegraphics[width=1\linewidth]{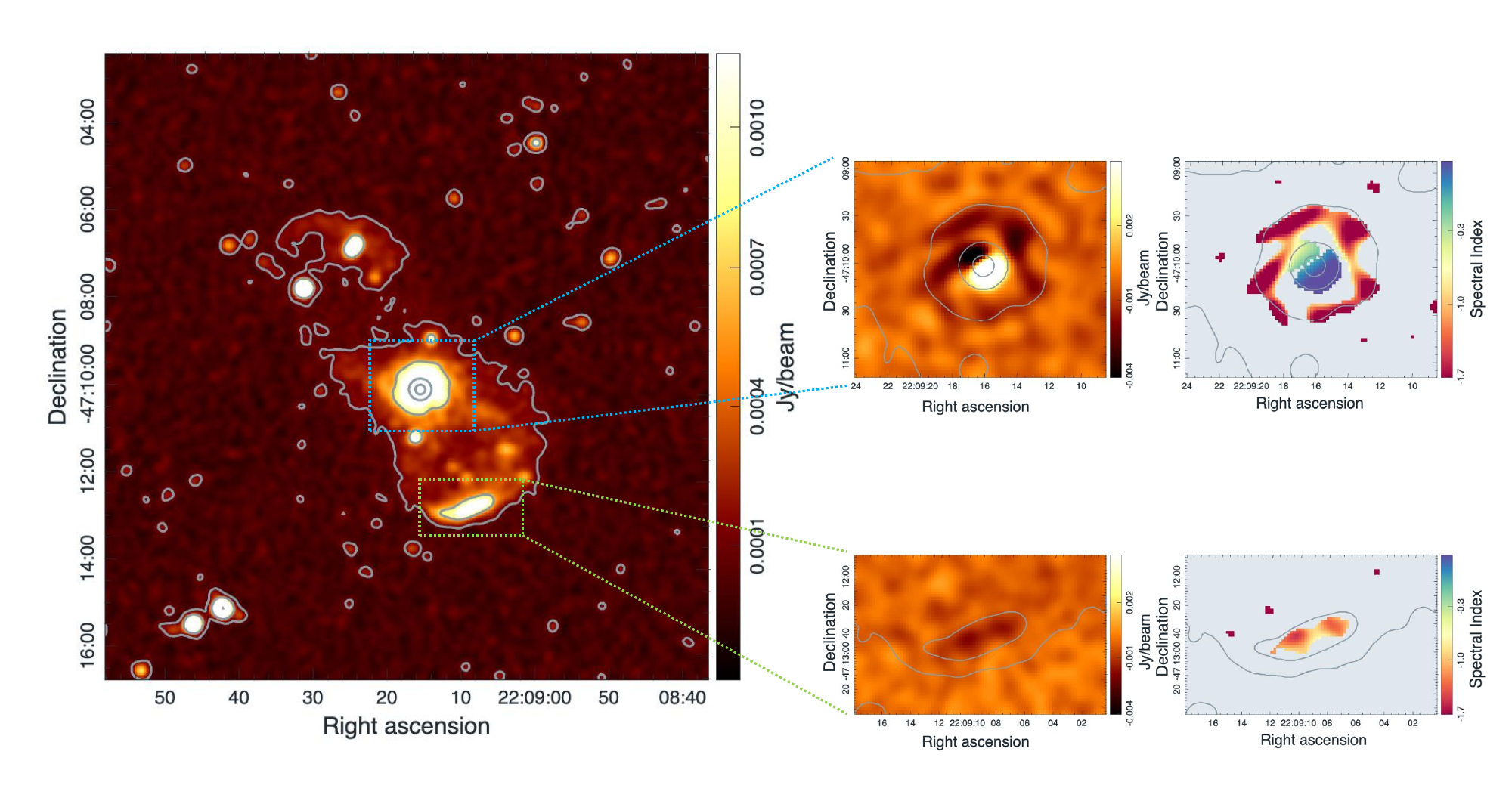}
    \caption{Spectral Index and Taylor~1 maps for \agn. The left image shows the larger Taylor~1 image, and the right side shows panel insets for the inner \ac{AGN} core and the patch of bright southern rim emission. The right-hand images show the uncut Taylor~1 image on the left and the cut spectral index (T1/T0) image on the right. All images have contours at levels 0.1, 0.7, 6, and 30\,mJy\,beam$^{-1}$ from the \ac{EMU} Taylor~0 image.}
    \label{fig:ASKAPspecindex}
\end{figure*}

\begin{figure*}[ht!]
    \centering
    \includegraphics[width=1\linewidth]{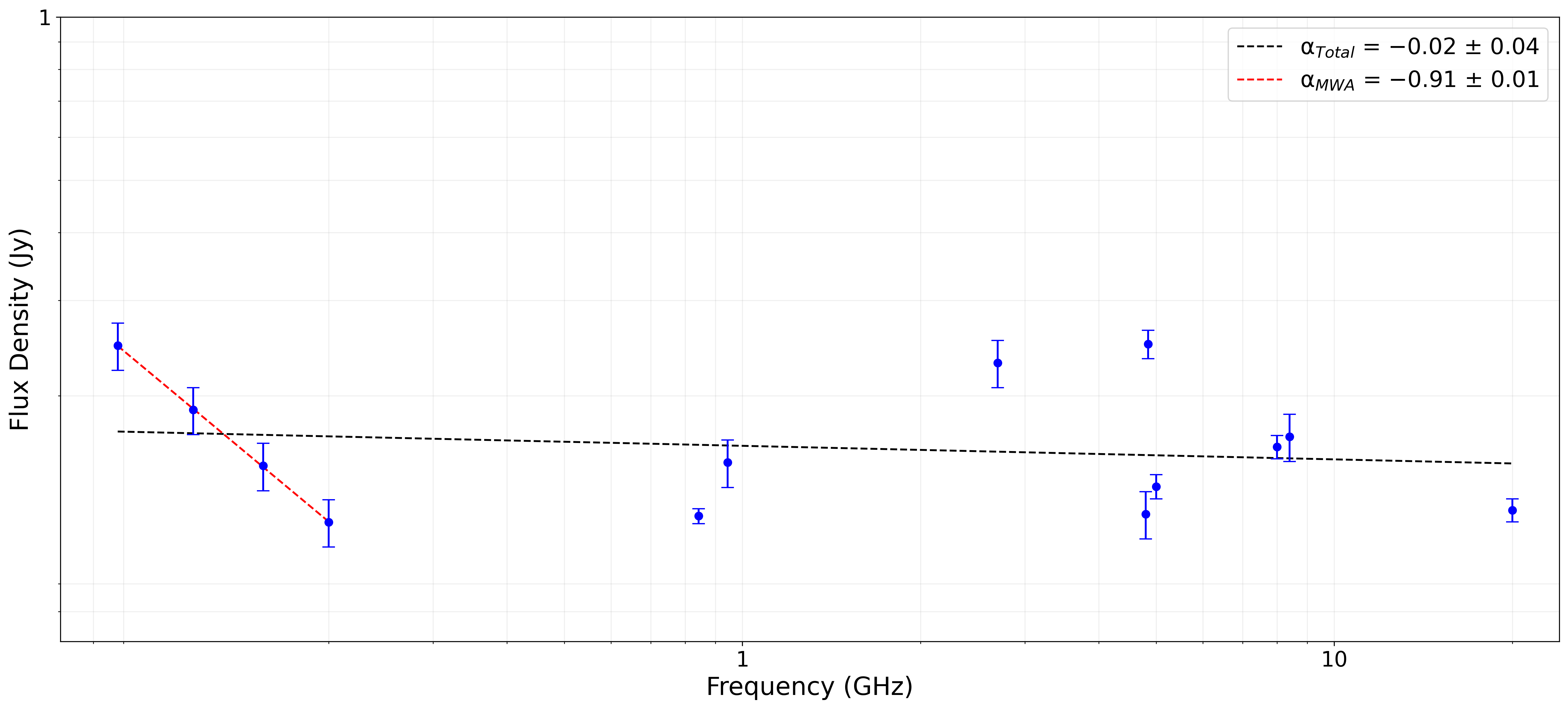}
    \caption{Spectral index plot for \agn\ using flux density measurements from Table~\ref{tab:fluxes}. The black dashed line shows the spectral index calculated from all flux density measurements in Table~\ref{tab:fluxes}, and the red dashed line shows the spectral index calculated from the \ac{MWA} points, as these \ac{MWA} observations were conducted at the same time.}
    \label{fig:specindex}
\end{figure*}

It is not possible to generate a spectral index map for the entire object due to the low surface brightness, but the spectral index can be estimated for the core and the brighter patch of southern radio emission (see Figure~\ref{fig:ASKAPspecindex}).

The Taylor~1 image shows a clear change between positive and negative coefficients within the two highest contour levels (see Figure~\ref{fig:ASKAPspecindex} delineate, top insets, left panel). These two highest contours show the inner, circular core, which appears as the brightest part of the radio-continuum image, indicating spectral variability within the inner core. A ring of negative T1 coefficients also surrounds this central core. This is also shown in the spectral index map (Figure~\ref{fig:ASKAPspecindex}, top insets, right panel), with a flat inner core ($\alpha\sim0$) and a much steeper ring structure ($\alpha\sim-1.7\,\text{to} -2.0$). The southern rim structure appears in the spectral index map with $\alpha\sim-1.0$, consistent with non-thermal radio emission.

This spectral index map was generated from the Taylor~0 and Taylor~1 images, and both were masked for values below 5$\sigma$ before creation. As the spectral index is calculated as $\alpha=\frac{\rm T1}{\rm T0}$, artificially steep values can arise if either term approaches the noise level, particularly if the denominator (T0) becomes small, leading to unstable ratios. Although we initially masked both Taylor images below the 5$\sigma$ level before generating the spectral index map, we further tested this by repeating the masking at the 7$\sigma$ and 10$\sigma$ levels.

The T0 image remains detected $\geq10\sigma$ across the entire core and ring region, precluding artificial inflation due to division by low signal-to-noise (S/N) T0 emission. In the T1 image, the core remains clearly detected at 10$\sigma$. These high S/N regions continue to show steep spectral indices ($\alpha\sim-2$), with the highest S/N pixels in T1 (S/N$\sim$27) reaching $\alpha\sim-2.2$. This confirms that the observed spectral steepening is not driven by noise fluctuations in T1. The negative T1 coefficients seen in the ring structure can arise from noise, deconvolution residuals, or intrinsic spectral structure. Given the high S/N demonstrated by these masking tests, we attribute the negative T1 values in the ring primarily to intrinsic spectral steepening, but acknowledge that the deconvolution residuals introduce systematic uncertainties which cannot be accounted for with the current data, making this indicative rather than definitive. The bottom rim structure is detected at a level above 5$\sigma$. It remains visible at 7$\sigma$ with a similar spectral index, but is not detected at 10$\sigma$, and so we treat this as a tentative detection.

A high-resolution multi-frequency radio spectral index map would be ideal, as it could help distinguish the outer large-scale lobes from the inner jet and characterise the different spectral ages. This would also help identify potentially distinct electron populations and provide insight into the timelines of previous \ac{AGN} activity episodes, for example, as in other analyses, including NGC\,2639 \citep{2023MNRAS.524.1615R,2022ApJ...940..168C}. This would require another suitably high-resolution radio observation at a different frequency, which is currently unavailable.

\begin{table}[ht!]
\caption{Integrated flux density measurements for \agn\ galaxy, including those measured in the current work and previous works with relevant references. Measurements indicated with $\dag$ are previous measurements where an uncertainty was not reported, and so an uncertainty of 10\% was assumed.}
    \centering
    \small
    \begin{tabular}{ccl}
    \hline\hline  
         $\nu$ (MHz) & $S\pm\Delta S$ (mJy) & Reference \\ 
    \hline
         88 & 247.6$\pm$24.8 & This work \\
         118 & 188.3$\pm$18.8 & This work \\
         155 & 148.5$\pm$14.9 & This work \\
         200 & 116.9$\pm$11.7 & This work \\
         843 & 120$\pm$3.8 & \citealt{2013yCat.8081....0M} \\
         944 & 150.6$\pm$15.1 & This work \\
         2700 & 230$\pm$23.0$^\dag$ & \citealt{1990PKS...C......0W}\\
         4800 & 121$\pm$ 12.1$^\dag$ & \citealt{2009MNRAS.397.2030H}\\
         4850 & 249$\pm$15 & \citealt{1994ApJS...91..111W}\\
         5000 & 136$\pm$7 & \citealt{2010MNRAS.402.2403M}\\
         8000 & 161$\pm$8 & \citealt{2010MNRAS.402.2403M}\\
         8400 & 168$\pm$16.8$^\dag$ & \citealt{2007ApJS..171...61H}\\
         20000 & 123$\pm$6 & \citealt{2010MNRAS.402.2403M} \\
         \hline\hline
    \end{tabular}
    \label{tab:fluxes}
\end{table}

\subsection{\agn\ Radio Polarisation}
\label{sec:polarisation}

From Figures~\ref{fig:polarisationWOptical2} and \ref{fig:polarisationWOptical3} (bottom), we can see a clear polarised intensity emission emanating from the central \agn\ \ac{SMBH} in the east ($\sim$90\arcsec\ or 10.5\,kpc) and, at a smaller scale ($\sim$30\arcsec\ or 3.5\,kpc), in the west direction. The possible jet axis is marked with the solid green line in Figure~\ref{fig:EMU} with a position angle of 75$^\circ$. This emission might represent \ac{AGN} jets that are unresolved in the larger Stokes~I image. So far, there are no reports of these jet structures. The inset in Figure~\ref{fig:EMU} shows the possible jet structure more clearly with log scaling, as the jets are faint compared to the core and are typically overpowered, making them difficult to see with linear scaling. 
What also stands out in Figure~\ref{fig:polarisationWOptical3} (bottom) is that the bright point source to the north exhibits B-fields suggestive of magnetic field compression, like something we see in “hotspots”, i.e. terminal shocked regions. Similarly, the “core” has B-fields in a certain direction. However, the vectors look more randomly oriented in other regions.

The core has about 0.5\% polarisation. A fractional polarisation level of $\sim$0.5\% is often comparable to typical residual instrumental polarisation leakage (see discussion in Section~2.1), but it can also be consistent with weak intrinsic core polarisation. For the diffuse polarised emission to the left (the eastern jet) of the bright core (\agn\ \ac{SMBH}), we estimate polarised emission at a level of $>50$\%, while the western jet is at a level of 5--7\%. The polarised structure in both jets is confined within the Stokes~$I$ contours in Figure~\ref{fig:polarisationWOptical2}, where the lowest level corresponds to $\sim$4$\sigma$, showing that the Stokes~$I$ emission is above the noise floor. This indicates that the high fractional polarisation is unlikely to be driven by division by low S/N intensity. Such high fractional polarisation (up to $\sim$50\%) is consistent with optically thin synchrotron emission in other resolved \ac{AGN} jet structures, where values of 20--50\% have been commonly observed~\citep{1984ARA&A..22..319B}. Also, it is not unusual to detect radio jets in face-on spirals, as shown in the case of NGC\,5938 \citep[][and reference therein]{2026PASA...43...35Z} and 6dFGS gJ035432.8--134008~\citep[J0354--1340;][]{2022A&A...662A..20V}.
We also note that the polarisation images of several Seyfert and \ac{LINER} galaxies (e.g., NGC\,4388, NGC\,4235, NGC\,3516, NGC\,3079) \citep{2025ApJ...989...40G} show how the Electric Vector Position Angle (EVPA) vectors (and thus the B-field vectors) vary smoothly from one region to the next, except in shock regions.
As the randomly oriented vectors in Figure~\ref{fig:polarisationWOptical3} are most likely real and significant (see Figure~\ref{fig:polarisationWOptical3}, right panels), the vector lengths proportional to fractional polarisation matter. This could either suggest a highly asymmetric medium causing differing amounts of Faraday rotation along the lobes, or the radio lobes being relic emission where “mixing” has happened between different media (thermal and non-thermal), disturbing the aligned magnetic fields seen in lobes that are currently being fed relativistic plasma from the central engine \citep{2020MNRAS.499..334S,2025ApJ...989...40G,2026ApJ..1000....3G} for NGC\,4151.

\section{DISCUSSION}
\label{discussion}

\citet{2020A&A...641A.151S} performed a multi-wavelength study (from X-ray to millimetres) of the \agn\ nucleus. They found that the central \ac{AGN} was too faint to significantly affect host-galaxy properties, such as star formation activity and molecular gas kinematics and distribution.

The X-ray study of \citet{2024ApJ...970...48S} found direct evidence for highly-ionised, fast hot winds (traced by ionised iron and other lines) launched from the hot accretion flow. They also report that the detection is consistent with winds at $\sim$10~keV that affect the pc-scale region. These highly ionised winds are consistent with the magnetised outflow scenario. This is important for \ac{AGN} feedback in \ac{LLAGN}.

The combination of hot winds, truncated disc signatures \citep{2022ApJ...926..209S}, and scarce molecular gas points to an accretion mode typical of low-accretion radiatively inefficient flows (e.g., hot inner flow / ADAF-like), rather than a luminous, thin disc. The hot wind detections suggest that even \ac{LLAGN} can launch energetically significant outflows that may regulate the gas supply \citep{2022ApJ...926..209S}. However, wide-angle winds emanating from the nucleus are ubiquitous among nearby Seyfert galaxies, such as NGC\,4945. Only relativistic winds are more common in Eddington-rate-accreting systems \citep{2025nfgs.book..277Y}.

As mentioned in Section~\ref{sec:spectra}, the \ac{FRED} X-ray flare suggests a long-term change in accretion (sudden increase then slow decay) -- models considered include a disc instability, a change in mass supply, or a delayed tidal disruption-like event, but the long timescale and properties are unusual for classic (stellar) \ac{TDE} \citep{2018MNRAS.475.1190Y}. 

In essence, previous studies suggest that \agn\ is a \ac{LLAGN}, and aside from the unusual southern H$\alpha$ feature (see Figure~\ref{fig:Halpha}), it is not an eventful object. However, our \ac{EMU} images show two new radio features, raising questions about their origin and the ``mix'' between the variable central \agn\ core on longer (year-scale) timescales, its radio jets, and/or two large-scale bidirectional galaxy hyperbubble features. These radio features do not obviously correlate with the observed optical spiral arm structure (see Figure~\ref{fig:polarisationWOptical2}, top panel), suggesting that they are not dominated by galactic star-forming disk emission. This is also consistent with \citet{2020MNRAS.499..334S}, who found that Seyferts display lobe/bubble-like radio morphologies not correlated with optical star-forming regions. They determined that excess radio emission in Seyferts is most likely due to jet-related emission and found that episodic \ac{AGN} activity is present in many of the Seyferts they analysed. This is consistent with the observed properties of \agn.

We now explore several possible scenarios for the origin of \agn\ and its extended radio emission found in our \ac{EMU} images.

\subsection{\agn\ as an \ac{ORC}}
\label{Discussion:ORC}

The two galaxy hyperbubbles seem morphologically similar to some newly suggested \ac{ORC}s \citep{2021PASA...38....3N,2022MNRAS.513.1300N,2021MNRAS.505L..11K,2025PASA...42..119S,2025A&A...702A.219T,2026PASA...43...43F}. 
However, the sizes of \agn\ hypothetical hyperbubbles differ by an order of magnitude, and therefore, we discard this as a realistic possibility.  

\subsection{\agn\ Southern Emission}
\label{Discussion:hyperbubble}

\citet{2001ApJ...546L..97H} identified a 19~kpc H$\alpha$ filament extending approximately 18.6~kpc south-west of the nucleus of \agn. They suggested that the filament was neutral gas photoionised by the active nucleus -- similar to that in the Seyfert~2 galaxy NGC\,5252 or possibly shock-ionised by a jet interacting with surrounding \HI, as suggested for the radio galaxy PKS~2240-41. 

Additionally, such features may be produced by an outflow or wind, interacting with the ambient \ac{ISM}, for example, in the case of NGC\,4258~\citep{2025ApJ...993..153F}. The galaxy ESO-G012~\citep{2026PASA...43...22K}, displays a similar ``hour-glass'' structure extending approximately 30\,kpc from the galaxy core, which is attributed to a bipolar outflow originating from the star-forming disk. A similar phenomenon may be generating the observed structure in \agn. 

\begin{figure*}[ht!]
    \centering
    \includegraphics[width=1\linewidth]{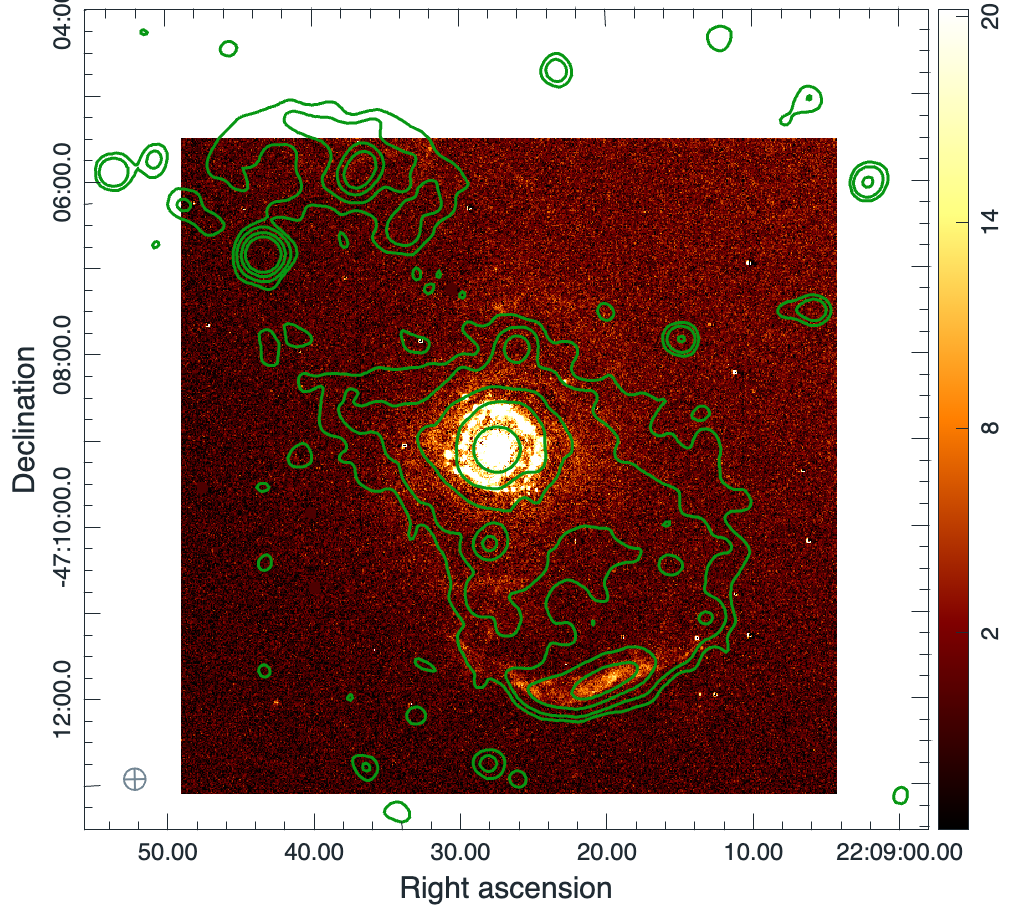}
    \caption{\agn\ H$\alpha$ image \citep{2001ApJ...546L..97H} overlaid with \ac{EMU} contours same as in Figure~\ref{fig:polarisationWOptical2}.}
    \label{fig:Halpha}
\end{figure*}

The \HI\ map of \citet{2001ApJ...546L..97H} reveals \agn\ to be a highly disturbed system, suggesting a past merger event, even though the \agn\ is not in a rich cluster environment \citep{1975gaun.book..557D}. This southern structure is visible in the \ac{EMU} radio-continuum image (see Figure~\ref{fig:Halpha}), but the current data are insufficient to distinguish between thermal and non-thermal emission. It shows no detectable radio polarisation, which may indicate thermal emission, but this could also reflect inherent sensitivity limits. The steep spectral index ($\alpha\sim-1.0$) is more indicative of a non-thermal origin, but this is a tentative detection calculated from the in-band \ac{ASKAP} spectral index. More radio observations are required to distinguish between these possibilities.

If the emission is thermal, this could suggest that the \agn\ southern feature is part of the galaxy's low-surface-brightness spiral arms or, alternatively, shock-ionised cones, as in NGC\,5252 \citep{1994AJ....107.1227W,2024ApJ...962..188W}. Some objects show radio emission that appears to follow the spiral arms of face-on spiral galaxies~\citep{2025MNRAS.543.3708B}. However, given our indicative steep spectral index for this region (see Section~\ref{sec:spectra}) as well as the morphological appearance of the rest of the radio-continuum emission from \agn, one would have difficulty reconciling \citet{2001ApJ...546L..97H} suggestions and the lack of other similar optical features across the multi-frequency comparison. 

If the emission were instead non-thermal, one might interpret this as the \ac{AGN} jet/outflow having changed direction over time. In this scenario, the steep spectral index may result from an older population of synchrotron-emitting electrons in the relic jet material, though further observations are required to test it. 

Alternatively, stellar processes such as supernovae or stellar winds can drive shocks producing similar structures. However, for the case of \agn, which is observed to have suppressed star formation and a molecular gas deficiency~\citep{2020A&A...641A.151S}, these are less likely drivers. \ac{AGN}-related activity, either through winds as evidenced by the highly-ionised fast outflows detected by \citep{2024ApJ...970...48S}, or through jet-\ac{ISM} interaction as suggested by the possible steep spectral index, therefore remains a more likely source of shock ionisation in this system.

Taken together, the large-scale ionised structures, evidence for galaxy-scale outflows, and extended radio emission suggest that \agn\ may host large, shock-heated gaseous structures (hyperbubbles). Although the present data do not uniquely distinguish this scenario from alternative interpretations, including \ac{AGN} photoionisation or jet-driven shocks.

\subsection{\agn\ as an FR-II Radio Galaxy}
\label{sec:Discussion:FR-II_Galaxy}

The radio brightening at the forward edges of the galaxy hyperbubbles resembles a bow shock, as seen in Fanaroff-Riley~Class~II (FR-II) radio galaxies \citep[see, e.g., Fig.~4 of][]{2025PASA...42..100Y}. 
One may consider the restarted jet as simulated by \citet[][their fig.~1e]{1991ApJ...369..308C}, where they show that when a restarted jet propagates through thermalised plasma from an earlier jet episode, it becomes overdense relative to the old lobe material, producing a protrusion at the leading edge of the lobe \citep[see also the case of III~Zw~2;][]{2021MNRAS.507..991S}.

The lack of definitive hotspots in our images (see Section~\ref{sec:polarisation}) suggests that the jets are either never collimated (because the whole galaxy is not in a gas-rich environment) or disrupted. While there is a difference in the \citet{2025PASA...42..100Y} results, with the jet being collimated and continuing, the basic bow-shock structure is similar, just wider in the case of \agn.
It is also possible that hotspots were previously present but later decayed or became disrupted due to their short lifetimes. Since the hotspot fading timescale is relatively short \citep[$\sim$10$^4$\,yr;][]{2001MNRAS.320..355H}, they may no longer be visible \citep[See also the case of PG~1004+130;][]{2023ApJ...958...71G}. However, we note that most recently \citet{2026A&A...710A.217B} suggested two radio-continuum point-like sources on the opposite side from the \ac{SMBH} as hotspots, which are also perpendicular to the here proposed inner jets of \agn. We note that those two suggested hotspots are not visible in our polarisation images (Figure~\ref{fig:polarisationWOptical3}).

\citet{2015MNRAS.454.1595K} analysed radio \ac{AGN} in spiral galaxies and found that objects like \agn\ do exist, but because the environments are different, the radio morphologies will typically be slightly different to those of the elliptical galaxies.
Partially synchrotron self-absorbed jets will show a declining (almost optically thin) power-law at low frequencies, then kick up again as shown in \citet[][fig.~12]{2025PASA...42..100Y}, especially the yellow 7\,kpc line in the left panel. This is similar in places to our Figure~\ref{fig:specindex}, except there is a lot of inconsistency in flux densities at nearby frequencies. This is most likely due to variability in the core of an \agn\ galaxy.

\subsection{Variability and accretion behaviour}
 \label{Discussion:variability}

Extended optical and X-ray fading in \agn\ has persisted for more than four decades, suggesting a major episodic accretion event that has slowly declined since the 1980s \citep{Yan2018}. While \ac{LLAGN} commonly show variability linked to stochastic inflow, the monotonic and prolonged decay of \agn\ stands out.

In the context of \agn, the coexistence of a low accretion rate ($L/L_{\mathrm{Edd}}\lesssim 10^{-3}$) and the absence of a classical, radiatively efficient disk implies that the jet is likely powered by a geometrically thick \ac{RIAF}.
In such systems, disk instabilities, changes in mass supply, or structural evolution of the accretion flow can produce long-term variability \citep{2009ApJ...699..626H,2026A&A...710A.217B}.

Variable obscuration by neutral gas may also contribute to the observed behaviour, particularly in the optical and X-ray bands. However, obscuration alone may not naturally explain the sustained, smooth decline observed over decades.

Thus, although disk-driven processes and obscuration provide the most straightforward explanation and serve as a natural baseline, the long-term coherence of the variability suggests that an additional mechanism may be influencing the system.

\subsection{Evidence for jet reorientation}
\label{Discussion:jet reorientation}

The combination of radio morphology and long-term variability suggests that the central engine of \agn\ may be undergoing structural evolution. In particular, the apparent misalignment between the inner polarised emission and the large-scale radio features points to a changing jet axis.

Such a reorientation could produce the observed ``sand-timer" or hourglass-like morphology, as successive outflows trace different directions over time. Additionally, it could explain the observed misalignment between the large-scale lobes and the inner, possibly newer, jet emission. Similar behaviour has been observed in \ac{AGN} where jet precession or reorientation leads to curved or multi-lobed radio structures (for example, see IRAS~06035--7102 in Velovi\'c~et al. (in prep.)).

Additional support for a dynamically evolving central region comes from the variability of broad emission lines, which show changes in both flux and profile over multi-year baselines. These variations are consistent with an evolving broad-line region (BLR), potentially linked to changes in the inner accretion flow or jet launching region.

Taken together, these observations suggest that the outflow in \agn\ is not fixed in orientation, but instead evolves over time. This naturally raises the question of the physical mechanisms underpinning such behaviour.

\subsection{Physical drivers of jet reorientation}
 \label{Discussion:drivers jet reorientation}

Several mechanisms may account for the apparent reorientation of the jet in \agn\, ranging from internal disk processes to more complex dynamical scenarios. The other plausible mechanisms for jet reorientation includes jet-ISM interactions \citep[e.g.][]{2006AJ....132..546G,2018MNRAS.476...80M}, 
accretion-disk warping, 
multiple \ac{AGN} episodes \citep[e.g., NGC\,2639;][]{2023MNRAS.524.1615R}, 
galaxy major and minor mergers and binary supermassive black holes, 
black hole spin reorientation following merger events, 
and Lense-Thirring precession. 

\subsubsection{Inner jets}
 \label{sec:inner_jets}

The polarisation emission from \agn\ as seen in Figures~\ref{fig:polarisationWOptical2} and \ref{fig:polarisationWOptical3} (marked with a green solid line in Figure~\ref{fig:EMU}) could represent ``inner jets''. Polarisation can be used as an indicator to distinguish between \ac{AGN} jet emission and that related to star formation, and higher fractional polarisation is not uncommon in Seyfert galaxies~\citep{2020MNRAS.499..334S}. If this emission does represent ``inner jets'', then this means that they start out in a direction not aligned with the lobes (galaxy hyperbubbles), but then reorient. This often happens due to buoyancy gradients \citep{2015MNRAS.454.1595K,2025PASA...42..100Y}. The relatively smaller size of the jets is also not unusual, as it has been observed that jets in Seyferts can be stunted, likely due to a close interplay between the jet and the surrounding material~\citep{2001A&A...369..467R,2020MNRAS.499..334S,2026PASA...43...35Z}.

We note that the south-western lobe appears brighter and more extended than the north-eastern counterpart. If this is the case, that these ``inner jets'' are the cause of the hyperbubbles, but not exactly aligned, then this would indicate that the south-western lobe is on the approaching side. This is slightly at odds with the observed polarisation. 
However, the receding jet would then have much larger fractional polarisation values and a greater extent of polarised emission ($>$50\% in the receding jet and 5--7\% in the approaching jet). This would represent an inverse Laing-Garrington effect~\citep{1988Natur.331..149L,1988Natur.331..147G}, wherein the receding jet appears more polarised. Several explanations have been proposed for this inverse effect, including strong environmental asymmetries, a massive starburst, or powerful superwinds. 
As \agn\ has been shown to exhibit suppressed star formation and a molecular gas deficiency~\citep{2020A&A...641A.151S}, we conclude that the massive starburst scenario is the least likely. This effect could be driven by environmental asymmetries, as supported by the general asymmetries seen in the larger bubble sizes, or by powerful superwinds, as supported by the highly ionised, fast, hot winds discovered by \citet{2024ApJ...970...48S}. Additionally, complexities in the internal magnetic field could cause significant differences in the observed polarisation. For example, one jet component might be more subject to mixing with the external environment or have a more tangled magnetic field~\citep{1988Natur.331..147G}. Complex magnetic field structures have been observed in Seyferts~\citep{2020MNRAS.499..334S}, which are seen to impact the observed fractional polarisation.

\subsubsection{Disk-driven mechanisms}
 \label{sec:Lense-Thirring}

In single black hole systems, jet precession can arise from misalignment between the black hole spin and the angular momentum of the accreting material. This can lead to warped or tilted accretion disks, with precession driven by Lense-Thirring torques arising from frame dragging by the spinning black hole.

In radiatively inefficient systems, such as \agn\, reduced viscous coupling allows such warps to persist over large scales. External torques from infalling gas or asymmetries in the accretion flow may further amplify these effects.

These processes provide a natural explanation for the gradual reorientation of jets without invoking additional components.

\subsubsection{Environmental perturbations}
\label{sec:env_pert}

Interactions with the surrounding medium may also influence jet direction. Asymmetric gas distributions, minor mergers, or residual disturbances from past interactions could perturb the accretion flow and alter the jet axis. While such effects can shape large-scale morphology, they are less likely to produce coherent, long-term precession of the central engine itself.

The large-scale misalignment of the inner jet may also be due to environmental interactions. If the jet is low-power or interacts with a dense \ac{ISM}, this may lead to jet destabilisation and reorientation, where the jet takes the path of least resistance or produces a decelerated, wind-like outflow~\citep{2006AJ....132..546G,2018MNRAS.476...80M}. This could result in the jets being oriented along the galaxy's minor axis or being correlated with the optical spiral arm structure in some way, which is not clearly observed (see Figures~\ref{fig:polarisationWOptical2} and \ref{fig:polarisationWOptical3}), but higher-resolution imaging of the jet would be required to confirm this.

\subsubsection{\agn\ as a \ac{TDE} in an binary \ac{SMBH} system}
\label{Discussion:SMBBH_and_TDE}


We explore the possibility that the observed long-lasting X-ray flare~\citep{2024ApJ...970...48S} could originate from a \ac{TDE} in a very eccentric \ac{SMBBH} system. The close pericentric approach of a secondary \ac{SMBH} may have perturbed a pre-existing disk \citep{Nixon2013}, triggering a delayed or prolonged \ac{TDE} that evolved into a \ac{RIAF}.

Interestingly, recent work on the radio-loud quasar 3C\,47 \citep{Marziani2025}, highlights similar evidence for binary SMBH–disk interactions, reinforcing the plausibility of such mechanisms in explaining extended flaring and asymmetric line structures in \ac{AGN}. Pericentric passages lead to episodic accretion flares that produce long-lived decays. The fallback and viscous timescales are sufficient to explain the 40-year decline in luminosity. The absence of renewed flaring supports a singular disruption event rather than continuous fueling.

Similar to 3C\,47, \agn\ shows an asymmetric line structure with double-peaked broad Balmer emission lines~\citep{Filipenko1984}, characteristic of \ac{SMBBH} candidates \citep{NguyenBogdanovic2016,Liu2016,Marziani2025a}. Changes in broad emission-line shapes can be used to constrain orbital parameters in some \ac{SMBBH} candidates \citep{Bon2012,Bon2016,Popovic2012,Li2019}. Similar changes in the broad \Halpha\ line have been detected in an orbiting spiral within the disk \citep{Schimoia2017}, which could indicate a binary scenario.

\agn\ exhibits the strongly double-peaked Balmer profiles seen in classical disk emitters, and the broad H$\alpha$ and H$\beta$ lines show measurable long-term variability in both flux and profile shape.
The Balmer-line wings display subtle asymmetries and centroid shifts over multi-year baselines, implying an evolving broad-line region (BLR).
Such variability is consistent with the \ac{SMBBH} scenario, although similar features can also be observed in single \ac{AGN} with variable accretion disks or BLR structure~\citep{2017ApJ...835..236S}.

Taken together, these results point toward a dynamic, gravitationally disturbed nucleus. This environment is naturally compatible with the presence of a secondary \ac{SMBH} on a mildly eccentric orbit that can gravitationally modulate inflow and trigger a large-scale \ac{TDE} or disk instability.
The observed four-decade decline in luminosity of \agn\ can thus be interpreted as the long-term viscous decay of such an event, consistent with models of binary-triggered TDEs \citep[e.g.,][]{Coughlin2017}.

The central compact radio structure of \agn, though relatively weak compared to classical radio-loud \ac{AGN}, shows possible morphological hints of mildly collimated nuclear outflows. 
Modest bending or reorientation of this jet-like feature may trace a slow precession of the launching axis, a phenomenon widely interpreted as evidence for \ac{SMBBH}s or warped accretion disks \citep{Caproni2004}.
Given the evidence for disturbed nuclear kinematics and inflowing spiral gas in \agn\ \citep{Schimoia2017}, the presence of a secondary \ac{SMBH} could exert sufficient torque to induce a long-term precession of the inner accretion flow.

A precessing jet in \agn, as seen in our \ac{EMU} images, could provide additional circumstantial evidence for a bound \ac{SMBH} pair modulating both the accretion disk and large-scale outflow. 

\subsubsection{Other scenarios}
 \label{sec:other_scenarios}

Given that the low accretion rate ($L/L_{\mathrm{Edd}}\lesssim 10^{-3}$) and the absence of a classical, radiatively efficient disk imply that the jet is likely powered by a geometrically thick \ac{RIAF}, even modest external torques from a companion \ac{SMBH} could induce significant disk warping, as the reduced viscous coupling allows large-scale tilt propagation \citep{Nixon2013}. 

A secondary black hole on an eccentric orbit could thus periodically realign the inner accretion flow and modulate both the jet direction and the accretion rate near pericenter passage. 
These modulations would manifest as long-term variability in optical, radio and X-ray emission. While current radio observations cover only a short timescale, previous observations over a four-decade timescale~\citep{Yan2018} are consistent with this scenario. 
In this picture, the optical emission-line region and the jet respond to the same underlying dynamical driver—the gravitational coupling of a binary \ac{SMBH} system with an evolving, partially truncated disk \citep{2022ApJ...926..209S}.

Moreover, if the secondary black hole periodically perturbs the inner accretion disk, it could lead to episodes of enhanced mass transfer or even partial tidal disruption of disk material, generating transient brightening followed by long-term decline.
The combined effect of jet-axis precession, episodic accretion, and fading disk emission could therefore reproduce the complex variability pattern observed in \agn\ without invoking external triggers such as major mergers.
This interpretation could be analogous to that proposed for the jetted quasar 3C\,47 \citep{Marziani2025}, in which the disk asymmetries and radio morphology have been attributed to binary-driven precession with a similar mass-ratio range ($q\sim0.5$).
In both cases, the precessing jet acts as an indirect tracer of the binary orbital configuration and provides observational access to sub-parsec-scale dynamics that would otherwise be unresolved.

All things considered, the binary-induced jet precession hypothesis for \agn\ unifies its observed radio properties, its long-term optical decline, and the variability of its double-peaked Balmer-line profiles within a single dynamical framework.
If confirmed through future \ac{VLBI} monitoring, line-shape variability, or future X-ray observations (e.g., {\it Chandra}), even subtle changes in jet orientation could provide powerful constraints on the orbital parameters and alignment history of the putative supermassive binary.
Such confirmation would make \agn\ an important nearby laboratory for studying the coupling among accretion-flow geometry, black-hole spin evolution, and binary dynamics in low-luminosity \ac{AGN}.

Perhaps the simplest explanation for this newly discovered \agn\ radio-continuum feature(s) is that it represents precessing and, at the same time, restarting jets that are originating from the \agn\ core as seen in a number of other galaxies \citep{2023ApJ...951..106B}. An \ac{SMBH} whose spin axis or accretion disk wobbles over time would cause its jets or outflows to sweep through space, creating a sand-timer or figure/number 8 morphological shape. This precession may be caused by a variety of factors, including the gravitational influence of a binary companion (\ac{SMBBH}) as suggested in Section~\ref{Discussion:SMBBH_and_TDE} \citep[also see:][]{Kun2014, Kun2024, Marziani2025}, a misalignment between the accretion disk and the \ac{SMBH} spin, or the spin of the \ac{SMBH} itself precessing \citep{Liska2018,Kun2023}.

\section{CONCLUSIONS}
\label{conclusion}

We present \ac{EMU} detection of several features in \agn\, some of which are reported here for the first time. The first and most obvious feature is the central \agn\ with its strong discrete radio source for which we estimate radio properties (see Sections~\ref{sec:flux} and \ref{sec:spectra}). The most striking feature appeared as two large-scale galaxy hyperbubbles, most likely due to emission from the central \ac{SMBH}, but their origin is unknown. The third feature is most likely directly related to the \ac{SMBH}, as two opposite radio jets emanating from it are clearly visible in the polarised intensity image but unresolved in the total intensity image. We also report an inner "ring-like" structure (see Figure~\ref{fig:specindex}, right) close to the central \agn\ core with a notably steeper radio spectral index, indicating a non-thermal nature different from the central core. 

We investigate several scenarios that could explain \agn's peculiar emission and conclude that the most likely scenario is that \agn\ hosts a \ac{SMBBH}, which produces restarting jets and central-core multi-frequency variability. Alternatively, the thermal hyperbubble, and especially the FR-II radio galaxy scenario, might also explain the radio-continuum properties.

\section*{Acknowledgments}

The Australia Telescope Compact Array is part of the Australia Telescope National Facility (\url{https://ror.org/05qajvd42}) which is funded by the Australian Government for operation as a National Facility managed by CSIRO. We acknowledge the Gomeroi people as the Traditional Owners of the Observatory site.
This scientific work uses data obtained from Inyarrimanha Ilgari Bundara, the CSIRO Murchison Radio-astronomy Observatory. We acknowledge the Wajarri Yamaji People as the Traditional Owners and native title holders of the Observatory site. CSIRO’s ASKAP radio telescope is part of the Australia Telescope National Facility (\url{https://ror.org/05qajvd42}). Operation of ASKAP is funded by the Australian Government with support from the National Collaborative Research Infrastructure Strategy. \ac{ASKAP} uses the resources of the Pawsey Supercomputing Research Centre. Establishment of ASKAP, Inyarrimanha Ilgari Bundara, the CSIRO Murchison Radio-astronomy Observatory and the Pawsey Supercomputing Research Centre are initiatives of the Australian Government, with support from the Government of Western Australia and the Science and Industry Endowment Fund.
This scientific work uses data obtained from Inyarrimanha Ilgari Bundara, the CSIRO Murchison Radio-astronomy Observatory. We acknowledge the Wajarri Yamaji People as the Traditional Owners and Native Title Holders of the observatory site. Support for the operation of the MWA is provided by the Australian Government (NCRIS), under a contract to Curtin University administered by Astronomy Australia Limited. We acknowledge the Pawsey Supercomputing Centre which is supported by the Western Australian and Australian Governments.
DU acknowledges the financial support provided by the Ministry of Science, Technological Development and Innovation of the Republic of Serbia through contract 451-03-136/2025-03/200104 and support through the joint project of the Serbian Academy of Sciences and Arts and Bulgarian Academy of Sciences ``Optical search for Galactic and extragalactic supernova remnants''. 
We thank E. Kammoun and E. Bon for their comments and suggestions, which improved this paper.
We also thank both referees for their comments and suggestions that significantly improved this paper.

\bibliographystyle{apsrev4-1}

\bibliography{NGC7213_OJA}

\end{document}